\documentclass[a4paper]{article}
\usepackage{ISCSLP2026}
\usepackage{ifthen}
\usepackage{multirow}
\usepackage{amsmath,amssymb,bm}
\usepackage{graphicx}
\usepackage{url} 
\newboolean{blind}
\setboolean{blind}{false}
\title{X-Pred MeanFlow for Streaming Token-to-Mel Speech Decoding}
\name{
    \ifthenelse{\boolean{blind}}
    {Anonymous to ISCSLP}
    {
    Hanke Xie$^1$,
    Xiaming Ren$^1$,
    Qirui Zhan$^1$,
    Jingbin Hu$^1$,
    Wenhao Li$^1$,
    Haoyu Zhang$^1$,
    Ruonan You$^1$,
    Chengyou Wang$^1$,
    Yunxiang Chen$^2$,
    Houdun Liu$^2$,
    Su Feng$^2$,
    Lei Xie$^{1,*}$
    }
}
\address{
  \ifthenelse{\boolean{blind}}{Anonymous to ISCSLP}
  {
  	$^1$Audio, Speech and Language Processing Group (ASLP@NPU), School of Software,\\
  	   Northwestern Polytechnical University, Xi'an, China\\
  $^2$Shenzhen Pimei Technology Co., Ltd.
  }
}
\email{
	\ifthenelse{\boolean{blind}}{Anonymous to ISCSLP}
	{hkxie@mail.nwpu.edu.cn, lxie@nwpu.edu.cn}
}

\begin{document}
\maketitle

\begin{abstract}
Recent advancements in discrete token-based speech generation have
highlighted the importance of efficient token-to-waveform synthesis in
streaming and dialogue scenarios. Flow-matching acoustic decoders
achieve high-quality token-to-mel generation, but their iterative
sampling requires multiple neural function evaluations, limiting
low-latency speech synthesis. MeanFlow reduces the sampling budget by
modeling the average velocity over a temporal interval, yet maintaining
high acoustic quality under extremely few-step token-to-mel generation
remains challenging. To address this challenge, we propose X-Pred MeanFlow, a few-step
streaming token-to-mel decoder that reparameterizes MeanFlow with
mel-space prediction. The decoder predicts a generalized mel field and
analytically derives the corresponding average velocity for sampling,
thereby preserving the MeanFlow formulation while providing a direct
acoustic prediction target. We further introduce layer-selective
block-wise attention to enable continuous chunk-wise generation with
bounded context. Experiments show that X-Pred MeanFlow improves
few-step token-to-mel synthesis over Direct-$u$ MeanFlow and supports
stable streaming generation. Speech samples are available.\footnote{\url{https://renxiaming.github.io/xpred-meanflow-stream-demo/}}

\end{abstract}

\begin{figure*}[t]
    \centering
    \includegraphics[width=0.98\textwidth]{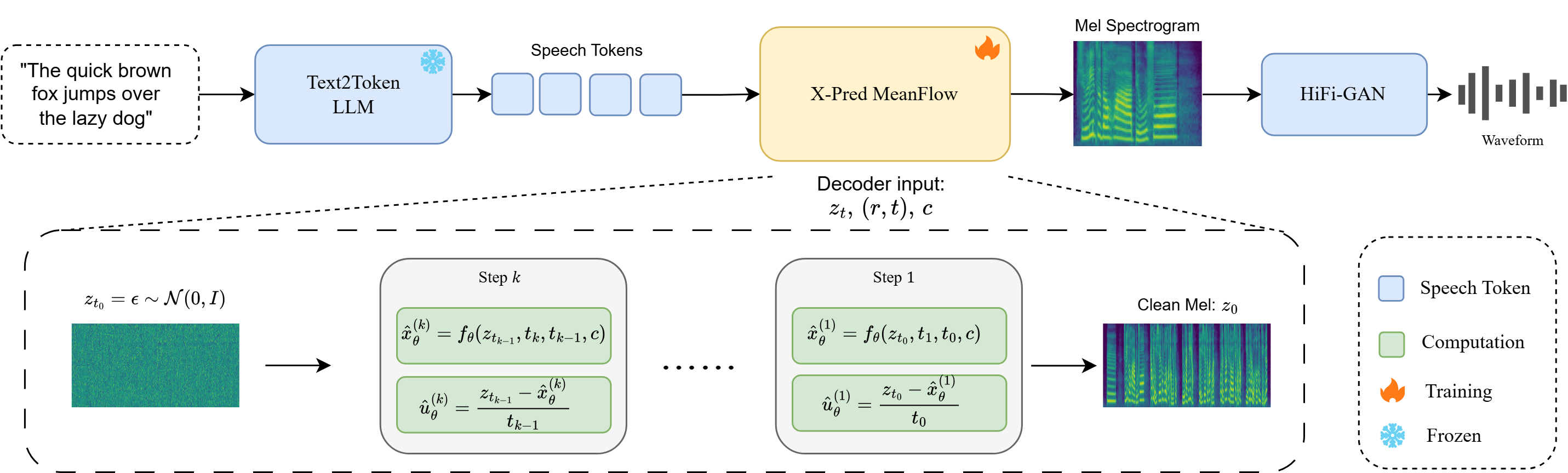}
    \caption{
        Overall framework and few-step inference process of the proposed
        X-Pred MeanFlow acoustic decoder.
        A frozen text-to-token LLM generates discrete speech tokens
        \(c\), which condition the trainable token-to-mel decoder.
        At each sampling step, the decoder directly predicts the clean
        mel-spectrogram \(\hat{x}_{\theta}\) and analytically recovers the
        mean velocity as
        \(\hat{u}_{\theta}=(z_t-\hat{x}_{\theta})/t\).
        The resulting mel-spectrogram \(\hat{x}\) is converted into waveform
        \(\hat{y}\) by a frozen HiFi-GAN vocoder.
        The snowflake symbol denotes frozen components.
    }
    \label{fig:xpred_overview}
    \vspace{-1mm}
\end{figure*}

\section{Introduction}

Recent progress in large language model (LLM)-based text-to-speech
(TTS) has shifted speech synthesis from direct regression of continuous
acoustic features~\cite{shen2018natural,ping2017deep,kim2020glow} to
sequence modeling over discrete speech representations
~\cite{wang2023vall,anastassiou2024seed,defossez2022high,
zeghidour2021soundstream,polyak2021speech}. Early neural codec language
models, represented by VALL-E and AudioLM, formulate speech generation
as conditional prediction of acoustic codec codes
~\cite{wang2023vall,borsos2023audiolm}. More recent systems increasingly
separate high-level speech content modeling from low-level acoustic
reconstruction~\cite{kharitonov2023speak,betker2023better,
casanova2024xtts,ye2025llasa,wang2025maskgct}. For example, CosyVoice
uses supervised semantic speech tokens for language modeling and a
conditional flow-matching decoder for acoustic generation
~\cite{cosyvoice,du2025cosyvoice}. This semantic-token pipeline provides
a compact interface for the language model, but it transfers the
recovery of speaker characteristics, prosody, and fine spectral
structure to the downstream acoustic decoder. Consequently,
token-to-acoustic decoding directly affects perceptual quality, the
computation required for each output packet, and whether speech can be
generated incrementally in streaming and dialogue scenarios.

Conditional flow matching has become an effective framework for
high-quality speech generation~\cite{DBLP:conf/iclr/LipmanCBNL23,
le2023voicebox,chen2024f5tts,zhu2025zipvoice,wang2025prosodyflow}. It
learns a continuous velocity field that transports samples from a prior
distribution to the target acoustic distribution and permits
non-autoregressive generation within each network evaluation. Inference,
however, still requires numerical integration of the learned field, and
its cost grows with the number of neural function evaluations (NFEs)
~\cite{song2020denoising,karras2022elucidating}. This cost is
particularly restrictive in streaming synthesis, where the computation
of every packet must finish within a fixed playback interval
~\cite{mehta2024matcha}. MeanFlow reduces the sampling budget by
modeling the average velocity over a temporal interval rather than only
the instantaneous velocity at a single time point~\cite{geng2025mean}.
Recent work further demonstrates the potential of average-velocity
modeling for few-step speech generation~\cite{wang2025intmeanflow}.
Nevertheless, preserving detailed acoustic structure under extremely
few-step token-to-mel generation remains challenging. Since the acoustic
decoder ultimately produces a mel-spectrogram, the output
parameterization of a speech-conditioned MeanFlow decoder becomes an
important design choice~\cite{li2026back}.

Sampling efficiency alone does not make an acoustic decoder streamable.
A decoder with global self-attention either depends on the complete
token sequence or attends to an increasingly long history during
incremental generation. At the other extreme, independently decoding
short chunks removes useful cross-chunk context and can produce
inconsistent acoustic transitions around chunk boundaries. CosyVoice~2
introduces chunk-aware causal flow matching to unify streaming and
non-streaming synthesis, while StreamFlow employs local block-wise
receptive fields across Diffusion Transformer layers to control
contextual dependencies~\cite{du2024cosyvoice2,guo2025streamflow}.
These studies establish structured receptive-field control as a key
component of streaming speech-token decoding. However, few-step
generation and bounded-context decoding are often optimized from
different perspectives: the former reduces the computation needed to
follow the generative trajectory, whereas the latter determines when a
chunk becomes available and how its computation scales with utterance
length.

A practical streaming token-to-mel decoder must therefore balance
several coupled objectives. It should retain acoustic quality when only
a few NFEs are available, keep its receptive field bounded so that
per-packet computation does not grow with the generated sequence, and
preserve sufficient neighboring context for smooth transitions between
chunks. These objectives introduce an inherent trade-off. Shorter
chunks reduce algorithmic waiting time but provide less acoustic
context, additional look-ahead supplies useful local information but
increases startup delay, and extra sampling steps refine the generated
mel-spectrogram at the expense of the real-time computation margin. The
central challenge is thus not only to accelerate flow sampling or to
constrain attention in isolation, but to coordinate the sampling budget
and the context budget within a unified acoustic decoder.

To address these limitations, we propose X-Pred MeanFlow, a few-step
streaming token-to-mel decoder. X-Pred predicts a generalized mel field
from an intermediate flow state and analytically derives the
corresponding average velocity required by MeanFlow sampling. This
mel-space parameterization preserves the MeanFlow identity and sampling
procedure while providing the decoder with a direct acoustic prediction
target. We further introduce layer-selective block-masked attention for
the Diffusion Transformer backbone~\cite{peebles2023dit}, where selected
layers exchange historical or bounded future context while most layers
focus on local acoustic refinement. This design bounds the receptive
field, makes the amount of look-ahead controllable, and supports both
strict streaming and bounded-look-ahead streaming. Experiments show that
X-Pred MeanFlow improves few-step token-to-mel quality over Direct-$u$
MeanFlow at both Small and Base model scales. Streaming evaluations
further demonstrate that layer-selective bounded context achieves a
favorable quality-efficiency trade-off and supports stable packet-level
generation within the real-time computation budget.

\section{Preliminaries}
\subsection{Conditional Flow Matching}

Conditional flow matching (CFM)~\cite{DBLP:conf/iclr/LipmanCBNL23,nguyen2023generative} learns a vector field to transport samples from a prior distribution $p_{\text{prior}}(\epsilon)$ to a data distribution $p_{\text{data}}(x)$. Given a data sample $x \sim p_{\text{data}}(x)$ and noise $\epsilon \sim \mathcal{N}(0, I)$, an optimal transport path is constructed as $z_t = (1-t)x + t\epsilon$, with the conditional velocity $v_t = dz_t/dt = \epsilon - x$. A neural network $f_\theta$ is trained to minimize the CFM objective:
\begin{equation}
\mathcal{L}_{\text{CFM}}(\theta)
= \mathbb{E}_{t,x,\epsilon}\left[
\left\lVert f_{\theta}(t,z_t)-v_t \right\rVert^2
\right].
\end{equation}
During inference, the model solves an ODE to recover $x$ from $\epsilon$, which typically requires multiple function evaluations and incurs non-trivial latency.

\subsection{MeanFlow}

To enable high-quality synthesis with only a single neural function evaluation (1-NFE), this work adopts mean flows~\cite{DBLP:journals/corr/MeanFlows,lu2026pixel_meanflow}. Given a time interval $[r, t]$, the average velocity along the ODE trajectory is defined as:
\begin{equation}
u(z_t, r, t) \triangleq \frac{1}{t-r} \int_r^t v(z_\tau, \tau)\, d\tau.
\end{equation}
Differentiating with respect to $t$ and rearranging yields the mean flows identity:
\begin{equation}
u(z_t, r, t) = v(z_t, t) - (t - r)\frac{d}{dt}u(z_t, r, t),
\end{equation}
where the total derivative is expanded via the Jacobian-vector product as $\frac{d}{dt}u = v(z_t, t)\partial_z u + \partial_t u$. Replacing the marginal velocity $v(z_t, t)$ with the conditional velocity $v_t = \epsilon - x$, the training target becomes $u_{\text{tgt}} = v_t - (t-r)(v_t \partial_z u_\theta + \partial_t u_\theta)$. The mean flows training objective is:
\begin{equation}
\mathcal{L}_{\text{MF}}(\theta)
= \mathbb{E}_{t,r,x,\epsilon}\left[
\left\lVert f_\theta(z_t,r,t)-\operatorname{sg}(u_{\text{tgt}}) \right\rVert^2
\right].
\end{equation}
where $\text{sg}(\cdot)$ denotes the stop-gradient operation. At $t = r$, this objective reduces to the standard CFM loss. During 1-NFE sampling, the clean sample is recovered as $z_0 = z_1 - f_\theta(z_1, 0, 1)$, where $z_1 = \epsilon \sim p_{\text{prior}}(\epsilon)$.

\section{Method}

\subsection{Overview}
Let $c=[c_1,\ldots,c_{T_c}]$ denote discrete speech tokens sampled at rate $f_c$, and let $x\in\mathbb{R}^{T_x\times D_m}$ denote the target mel-spectrogram sampled at rate $f_m$. We define the temporal-rate ratio as $\rho=f_m/f_c$, such that each token is aligned with $\rho$ mel frames. Our study focuses on the conditional acoustic mapping $c\rightarrow x$; the upstream token generator and downstream vocoder are fixed and excluded from model optimization.

Figure~\ref{fig:xpred_overview} presents the overall framework. The decoder uses a DiT backbone and combines two complementary designs. X-Pred MeanFlow reparameterizes the MeanFlow predictor through a mel-space output to improve generation under a small number of neural function evaluations. Layer-selective block-masked attention restricts the acoustic context to a bounded history and optional look-ahead, enabling chunk-wise inference with duration-independent per-packet computation. The two components respectively control the sampling budget and the context budget of streaming token-to-mel decoding.

\begin{figure}[t]
    \centering
    \includegraphics[width=\linewidth]{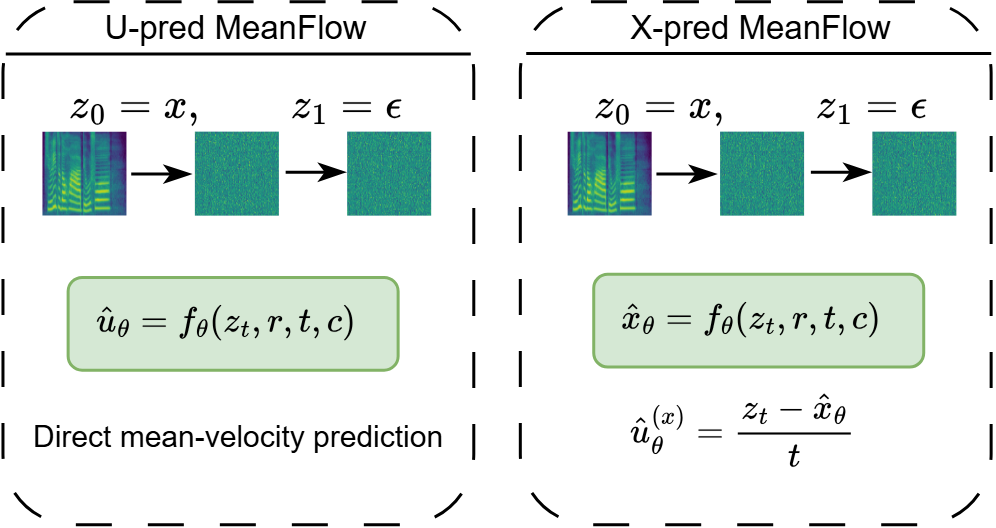}
    \caption{Comparison between direct-$u$ MeanFlow and X-Pred MeanFlow. Both use the same MeanFlow target and sampling rule. Direct-$u$ MeanFlow predicts the average velocity, whereas X-Pred MeanFlow predicts the generalized mel field $\hat{x}_{\theta}$ and obtains $\hat{u}_{\theta}=(z_t-\hat{x}_{\theta})/t$ analytically. When $r=0$, $\hat{x}_{\theta}$ is the predicted clean mel endpoint.}
    \label{fig:xpred_parameterization}
\end{figure}
\subsection{X-Pred MeanFlow}
\label{sec:xpred_meanflow}

A conventional MeanFlow decoder directly parameterizes the average
velocity $u(z_t,r,t,c)$. In speech token-to-mel decoding, however,
the quantity of ultimate interest is the mel-spectrogram itself rather
than the velocity field. We therefore reparameterize the prediction in
the mel space by inducing a data-space field from the same average
velocity. Specifically, we define the generalized mel field as
\begin{equation}
    x^{\mathrm{MF}}(z_t,r,t,c)
    \triangleq z_t - t\,u(z_t,r,t,c).
    \label{eq:generalized_mel_field}
\end{equation}
This quantity is indexed by both interval endpoints $(r,t)$ and is
therefore not always identical to the clean sample $x$, but it has a
direct interpretation in the mel space. When $r=0$, the average
velocity satisfies $u(z_t,0,t,c)=(z_t-z_0)/t$, and
Eq.~\eqref{eq:generalized_mel_field} reduces to
$x^{\mathrm{MF}}=z_0$, i.e., the clean mel endpoint. When $r=t$, the
average velocity collapses to the instantaneous velocity, and the
induced field becomes the standard endpoint-prediction target under the
conditional interpolation path. For $0<r<t$, it represents an
interval-conditioned, denoised mel-space quantity.

Our decoder predicts this generalized mel field directly:
\begin{equation}
    \hat{x}_\theta = f_\theta(z_t,r,t,c),
    \label{eq:xpred_output}
\end{equation}
and analytically recovers the corresponding average velocity as
\begin{equation}
    \hat{u}_\theta(z_t,r,t,c)=\frac{z_t-\hat{x}_\theta}{t}, \qquad t>0.
    \label{eq:xpred_to_u}
\end{equation}
This is a one-to-one reparameterization: the network output is expressed
in the mel space, while the MeanFlow dynamics are still carried out in
the velocity space. The denominator is $t$, rather than $t-r$, because
the induced mel field is defined relative to the endpoint at time zero.

Let $v_t$ denote the conditional velocity along the interpolation path.
The material derivative of the predicted average velocity is
\begin{equation}
    D_t\hat{u}_\theta=\partial_t\hat{u}_\theta+\partial_z\hat{u}_\theta\cdot v_t,
    \label{eq:material_derivative}
\end{equation}
where the Jacobian-vector product is computed by automatic
differentiation. The MeanFlow target is then written as
\begin{equation}
    u^{\mathrm{tgt}}_t = v_t-(t-r)D_t\hat{u}_\theta,
    \label{eq:meanflow_target_xpred}
\end{equation}
and the training objective becomes
\begin{equation}
    \mathcal{L}_{\mathrm{XMF}}
    = \mathbb{E}_{x,\epsilon,r,t,c}
    \left[\left\|\hat{u}_\theta-\mathrm{sg}(u^{\mathrm{tgt}}_t)\right\|_2^2\right].
    \label{eq:xpred_meanflow_loss}
\end{equation}
Therefore, the optimization target remains exactly in the MeanFlow
velocity space, while the network output is parameterized in the mel
space. At the boundary $r=t$, the correction term vanishes and
\begin{equation}
    \hat{u}_\theta-v_t = \frac{x-\hat{x}_\theta}{t},
    \label{eq:xpred_boundary_relation}
\end{equation}
which recovers a $t^{-2}$-weighted endpoint-prediction form.

During inference, let $1=t_0>t_1>\cdots>t_K=0$. Starting from
$z_{t_0}\sim\mathcal{N}(0,I)$, we update
\begin{equation}
    z_{t_k}=z_{t_{k-1}}-(t_{k-1}-t_k)\,
    \hat{u}_\theta(z_{t_{k-1}},t_k,t_{k-1},c).
\end{equation}
At each step, the decoder first predicts the generalized mel field and
then analytically converts it into the MeanFlow update direction through
Eq.~\eqref{eq:xpred_to_u}. In the 1-NFE case, the update reduces to
$z_0=\hat{x}_\theta(z_1,0,1,c)$, so the model output is exactly the
predicted clean mel endpoint.

\begin{figure}[t]
    \centering
    \includegraphics[width=\linewidth]{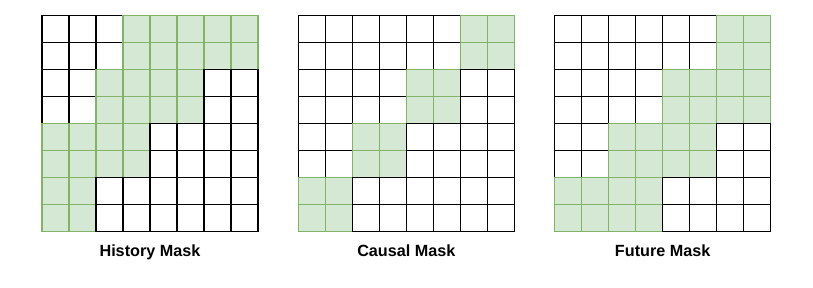}
    \caption{Layer-selective block-masked attention. Selected DiT layers use history, causal, or future masks to form a bounded receptive field for chunk-wise streaming inference.}
    \label{fig:block_attention}
\end{figure}

\subsection{Layer-Selective Block-Masked Attention}
\label{sec:block_attention}
Full-context self-attention requires the complete acoustic sequence and makes streaming cost grow with utterance length. We instead divide the token and mel sequences into temporally aligned blocks. Let each token block contain $C$ tokens. Its corresponding mel block contains $B=\rho C$ frames, and the $k$-th mel block and frame-to-block mapping are
\begin{equation}
    X_k=x_{kB:(k+1)B-1},
    \qquad
    q(i)=\left\lfloor\frac{i}{B}\right\rfloor.
    \label{eq:block_alignment}
\end{equation}
The aligned token block is used as the conditioning input for the same acoustic interval.

At DiT layer $\ell$, we assign a past-block budget $P_{\ell}$ and a future-block budget $F_{\ell}$. The additive attention mask is
\begin{equation}
M^{(\ell)}_{ij}=
\begin{cases}
0,
& -P_{\ell}\leq q(j)-q(i)\leq F_{\ell},\\
-\infty,
& \text{otherwise},
\end{cases}
\label{eq:layer_selective_mask}
\end{equation}
and is added to the self-attention logits before softmax. The setting $(P_{\ell},F_{\ell})=(0,0)$ restricts attention to the current block. A positive $P_{\ell}$ exposes historical blocks, while a positive $F_{\ell}$ introduces bounded future context. These budgets are assigned selectively across the DiT depth, allowing most layers to refine local acoustic structure while a subset exchanges information across block boundaries. Figure~\ref{fig:block_attention} illustrates the resulting local, history, and future masks.

The directly visible acoustic span at layer $\ell$ is
\begin{equation}
    R^{\mathrm{dir}}_{\ell}
    =(P_{\ell}+1+F_{\ell})B.
    \label{eq:direct_receptive_field}
\end{equation}
Because information can propagate through multiple masked layers, this quantity does not by itself bound the complete network receptive field. We therefore perform packet-level inference within a fixed sliding window
\begin{equation}
    \mathcal{W}_k
    =\{X_{k-H},\ldots,X_k,\ldots,X_{k+F}\},
    \label{eq:streaming_window}
\end{equation}
where $H$ and $F$ are the available historical and future blocks. Frames outside $\mathcal{W}_k$ are never presented to the decoder, which provides a strict upper bound on context, memory, and per-packet computation.

\section{Experiments}

\subsection{Experimental Setup}

\noindent\textbf{Training Setup.} We train the acoustic decoder on the Emilia corpus~\cite{he2024emilia}. 
Speech is represented by S3Tokenizer v2 tokens with a CosyVoice2 BPE 
vocabulary of 6,563 entries at 25~Hz. The target representation is an 
80-bin mel-spectrogram extracted at 16~kHz with a 160-sample hop and a 
1,024-point FFT. The same fixed HiFi-GAN vocoder is used for all systems. 
We use AdamW with a peak learning rate of $7.5\times10^{-5}$, 20k 
warmup updates, gradient accumulation of 2, and a frame-based batch size 
of 2,000 per GPU.

\noindent\textbf{Model Details.} We denote the 16-layer, 159M-parameter DiT with hidden size 768 
and 16 attention heads as \textit{Small}, and the 22-layer, 
341M-parameter DiT with hidden size 1,024 as \textit{Base}. Within each 
model scale, Direct-$u$ and X-Pred use the same backbone, data, 
block-mask configuration, optimizer, and 2.0M-update training budget. 
X-Pred uses a CFG-aware training target with a guidance scale of 2.0, 
while all reported inference is performed with $\mathrm{CFG}=0$.

\noindent\textbf{Evaluation Metrics.} Evaluation uses 100 randomly selected sentences from 
unseen speakers. We report UTMOS~\cite{saeki2022utmos}, word error rate 
(WER), and speaker similarity (SIM). WER and SIM are computed using the 
official SEED-TTS evaluation toolkit.\footnote{https://github.com/BytedanceSpeech/seed-tts-eval}

\subsection{Few-Step Acoustic Quality}

Table~\ref{tab:few_step_quality} compares the generation quality between Direct-$u$ and X-Pred. Under stringent low-step constraints, X-Pred consistently outperforms the Direct-$u$ baseline across all metrics for both Small and Base configurations, confirming that the mel-space parameterization effectively accelerates high-quality acoustic convergence. Notably, the Small X-Pred model achieves higher overall speech quality than the Base Direct-$u$ variant at identical steps, while maintaining highly competitive similarity and intelligibility, demonstrating significant parameter efficiency. While extending the inference to 10 NFEs provides a reliable quality upper bound, the foundational performance advantages of X-Pred are already firmly established within the 2 to 3 NFE range, making it highly optimal for demanding streaming applications.

\begin{table}[t]
    \caption{
        Few-step token-to-mel generation quality (2.0M updates, $\mathrm{CFG}=0$). 
        Bold indicates the better result; 
        10-NFE serves as a quality reference.
    }
    \label{tab:few_step_quality}
    \centering
    \scriptsize
    \setlength{\tabcolsep}{4.0pt}
    \begin{tabular}{@{}lcccc@{}}
        \toprule
        \textbf{Model} &
        \textbf{NFE} &
        \textbf{UTMOS $\uparrow$} &
        \textbf{SIM $\uparrow$} &
        \textbf{WER (\%) $\downarrow$} \\
        \midrule

        \multicolumn{5}{c}{\textit{Small}} \\
        \midrule 
        Direct-$u$    & 2  & 2.910 & 0.638 & 8.19 \\
        Direct-$u$    & 3  & 3.051 & 0.649 & 7.79 \\
        \addlinespace
        X-Pred (Ours) & 2  & \textbf{3.189} & \textbf{0.661} & \textbf{6.89} \\
        X-Pred (Ours) & 3  & \textbf{3.310} & \textbf{0.670} & \textbf{6.28} \\
        X-Pred (Ours) & 10 & 3.464 & 0.697 & 5.54 \\

        \midrule

        \multicolumn{5}{c}{\textit{Base}} \\
        \midrule 
        Direct-$u$    & 2  & 2.867 & 0.669 & 6.58 \\
        Direct-$u$    & 3  & 3.076 & 0.671 & 6.30 \\
        \addlinespace
        X-Pred (Ours) & 2  & \textbf{3.247} & \textbf{0.678} & \textbf{6.23} \\
        X-Pred (Ours) & 3  & \textbf{3.382} & \textbf{0.684} & \textbf{5.89} \\
        \bottomrule
    \end{tabular}
\end{table}

\subsection{Streaming Quality and Runtime}

Table~\ref{tab:streaming_joint} evaluates the impact of various context schedules on generation quality and system efficiency. While independently decoding short blocks under the Local configuration yields the lowest computational overhead, it severely degrades acoustic quality and boundary continuity, indicating that isolated generation is insufficient for continuous speech. Applying a uniform bounded mask across all layers recovers a significant portion of this quality loss but introduces high processing latency. By restricting cross-block communication to specific layers and incorporating bounded future look-ahead, the LS-Bounded approach achieves an optimal balance. Compared to the Uniform baseline, LS-Bounded simultaneously improves overall speech quality and boundary smoothness while effectively reducing the real-time factor and estimated startup latency. Extensive runtime evaluations over hundreds of seconds of audio confirm the stability of this approach; at both 2 and 3 NFEs, LS-Bounded operates comfortably within the strict playback budget, exhibiting zero deadline misses and maintaining highly stable steady-packet latencies for robust real-time acoustic decoding..

\begin{table}[t]
    \caption{
Streaming quality and runtime under various context schedules.
Full represents the non-streaming B-CMOS reference.
LS denotes layer-selective attention.
First/Startup indicates initial packet computation and estimated system startup latency. Bold text highlights the optimal streaming performance at 3 NFEs.
    }
    \label{tab:streaming_joint}
    \centering
    \scriptsize
    \setlength{\tabcolsep}{2.2pt}
    \begin{tabular}{@{}lccccc@{}}
        \toprule
        \textbf{Config.} &
        \textbf{NFE} &
        \textbf{UTMOS $\uparrow$} &
        \textbf{B-CMOS $\uparrow$} &
        \textbf{RTF $\downarrow$} &
        \shortstack{\textbf{First/Startup}\\
        \textbf{(ms) $\downarrow$}} \\
        \midrule
        Full       & 3 & 3.356 & 0.00  & --    & -- \\
        Local      & 3 & 3.084 & -0.42 & 0.352 & 98.0/338.0 \\
        Uniform    & 3 & 3.273 & -0.15 & 0.618 & 160.0/640.0 \\
        LS-Strict  & 3 & 3.238 & -0.23 & 0.446 & 118.0/358.0 \\
        \midrule
        LS-Bounded & 2 & 3.189 & -0.10 & 0.401 & 106.9/586.9 \\
        LS-Bounded & 3 & \textbf{3.310} & \textbf{-0.07}
                         & 0.543 & 141.0/621.0 \\
        \bottomrule
    \end{tabular}
\end{table}

\section{Conclusion}
We present X-Pred MeanFlow, a few-step streaming token-to-mel decoder
with mel-space MeanFlow parameterization and layer-selective
block-masked attention. X-Pred predicts a generalized mel field and
analytically derives the corresponding average velocity, preserving the
MeanFlow sampler while giving the decoder a direct acoustic prediction
target. Experiments show that X-Pred consistently improves few-step
UTMOS, SIM, and WER over Direct-$u$ MeanFlow at both Small and Base
scales. The Small X-Pred decoder also outperforms the Base Direct-$u$
model in low-NFE naturalness while maintaining comparable speaker
similarity and intelligibility, showing favorable parameter efficiency.
For streaming inference, layer-selective bounded context provides the
best quality-efficiency trade-off among the evaluated context schedules,
and packet-level measurements confirm stable real-time acoustic decoding
under both 2-NFE and 3-NFE settings.


\bibliographystyle{IEEEtran}
\bibliography{mybib}

\end{document}